\documentclass[12pt]{article}
\usepackage{graphicx}

\begin{document}

\baselineskip=18pt plus 1pt minus 1pt

\begin{center}{\large \bf 
$\gamma$-rigid solution of the Bohr Hamiltonian for $\gamma=30^{\rm o}$
compared to the E(5) critical point symmetry
} 

\bigskip\bigskip

{Dennis Bonatsos$^{a}$\footnote{e-mail: bonat@inp.demokritos.gr},
D. Lenis$^{a}$\footnote{e-mail: lenis@inp.demokritos.gr}, 
D. Petrellis$^{a}$\footnote{e-mail: petrellis@inp.demokritos.gr}, 
P. A. Terziev$^{b}$\footnote{e-mail: terziev@inrne.bas.bg} },
I. Yigitoglu$^{a,c}$\footnote{e-mail: yigitoglu@istanbul.edu.tr}
\bigskip

{$^{a}$ Institute of Nuclear Physics, N.C.S.R.
``Demokritos''}

{GR-15310 Aghia Paraskevi, Attiki, Greece}

{$^{b}$ Institute for Nuclear Research and Nuclear Energy, Bulgarian
Academy of Sciences }

{72 Tzarigrad Road, BG-1784 Sofia, Bulgaria}

{$^{c}$ Hasan Ali Yucel Faculty of Education, Istanbul University}

{TR-34470 Beyazit, Istanbul, Turkey} 

\end{center}

\bigskip\bigskip
\centerline{\bf Abstract} \medskip

A $\gamma$-rigid solution of the Bohr Hamiltonian for $\gamma=30^{\rm o}$
is derived, its ground state band being related to the second order Casimir
operator of the Euclidean algebra E(4). Parameter-free (up to overall scale 
factors) predictions for spectra and B(E2) transition rates are in close 
agreement to the E(5) critical point symmetry, as well as to experimental 
data in the Xe region around $A=130$. 

\bigskip 
{\bf 1. Introduction}

The E(5) \cite{IacE5} and X(5) \cite{IacX5} critical point symmetries, 
describing shape phase transitions from vibrational [U(5)] to 
$\gamma$-unstable [SO(6)] and vibrational to prolate deformed [SU(3)] 
nuclei respectively, have attracted recently much attention, since 
supporting experimental evidence is increasing 
\cite{CZE5,ClarkE5,CZX5,ClarkX5}. The E(5) model is obtained as an exact 
solution of the Bohr Hamiltonian \cite{Bohr} for $\gamma$-independent 
potentials 
\cite{IacE5}, while the X(5) model is obtained as an approximate solution 
for $\gamma \approx 0^{\rm o}$ \cite{IacX5}. Another approximate solution, 
with $\gamma \approx 30^{\rm o}$, called Z(5), has also been obtained 
\cite{Z5}. In all these cases, five degrees of freedom (the collective 
variables $\beta$, $\gamma$, and the three Euler angles) are taken into 
account.

In the present work we derive an exact solution of the Bohr Hamiltonian 
for $\gamma=30^{\rm o}$, by ``freezing'' $\gamma$ (as in Ref. \cite{DavCha})
to this value and taking 
into account only four degrees of freedom ($\beta$ and the Euler angles). 
In accordance to previous terminology, this solution will be called Z(4).  
It turns out that the Z(4) spectra and B(E2) transition rates are quite 
similar to the E(5) ones, while in parallel the ground state band of Z(4) 
is related to the Euclidean algebra E(4), thus offering the first clue of 
connection between critical point symmetries and Lie algebraic symmetries,
in addition to the case of the E(5) model \cite{IacE5}.   
Experimental examples of Z(4) seem to appear in the Xe region around $A=130$. 

The Z(4) solution will be introduced in Section 2 and its ground state 
band will be related to E(4) in Section 3. Numerical results and comparisons 
to E(5) and experiment will be given in Section 4, while discussion of the 
present results and plans for further work will appear in Section 5. 

{\bf 2. The Z(4) model} 

In the model of Davydov and Chaban \cite{DavCha} it is assumed that the 
nucleus is rigid with respect to $\gamma$-vibrations. Then the Hamiltonian 
depends on four variables ($\beta,\theta_i$) and has the form \cite{DavCha} 
\begin{equation}\label{eq:e41} 
H = -\frac{\hbar^2}{2B}\Biggl[\frac{1}{\beta^3}\frac{\partial}{\partial\beta}
\beta^3\frac{\partial}{\partial\beta} - \frac{1}{4\beta^2}
\sum_{k=1}^{3}\frac{Q_{k}^2}{\sin^2(\gamma-\frac{2\pi}{3}k)}\Biggr]
+ U(\beta),
\end{equation}
where $\beta$ and $\gamma$ are the usual collective coordinates \cite{Bohr}, 
while $Q_k$ ($k=1$, 2, 3) are the components of angular momentum and $B$ is 
the mass parameter. In this Hamiltonian $\gamma$ is treated as a parameter and 
not as a variable. The kinetic energy term of Eq. (\ref{eq:e41}) is different 
from the one appearing in the E(5) and X(5) models, because of the different 
number of degrees of freedom treated in each case (four in the former case, 
five in the latter). 
 
Introducing \cite{IacE5} reduced energies $\epsilon= (2 B/\hbar^2) E$ and 
reduced 
potentials $u=(2B/\hbar^2) U$, and considering a wave function of the form 
$\Psi(\beta,\theta_i)=\phi(\beta)\psi(\theta_i)$, where $\theta_i$ (
$i=1$, 2, 3) are the Euler angles, 
separation of variables 
leads to two equations
\begin{equation}\label{eq:e42} 
\Biggl[\frac{1}{\beta^3}\frac{\partial}{\partial\beta}
\beta^3\frac{\partial}{\partial\beta} - \frac{\lambda}{\beta^2}
+(\epsilon-u(\beta))\Biggr] \phi(\beta) = 0, 
\end{equation}
\begin{equation}\label{eq:e43} 
\Biggl[\frac{1}{4}\sum_{k=1}^{3}\frac{Q_{k}^2}{\sin^2(\gamma-\frac{2\pi}{3}k)}
-\lambda \Biggr]\psi(\theta_i) = 0.
\end{equation}

In the case of $\gamma=\pi/6$, the last equation takes the form 
\begin{equation}\label{eq:e44}
\Biggl[\frac{1}{4}(Q_1^2 + 4Q_2^2 + 4Q_3^2)-\lambda \Biggr]\psi(\theta_i) = 0.
\end{equation}
This equation has been solved by Meyer-ter-Vehn \cite{MtVNPA}, the 
eigenfunctions being  
\begin{equation}\label{eq:e45}
\psi(\theta_i)=\psi^L_{\mu,\alpha}(\theta_i) =
\sqrt{\frac{2L+1}{16\pi^2(1+\delta_{\alpha,0})}}\,
\Bigl[{\cal D}^{(L)}_{\mu,\alpha}(\theta_i) 
+ (-1)^L {\cal D}^{(L)}_{\mu,-\alpha}(\theta_i)\Bigr]
\end{equation}
with 
\begin{equation}\label{eq:e46}
\lambda = \lambda_{L,\alpha} = L(L+1) -\frac{3}{4}\,\alpha^2, 
\end{equation}
where ${\cal D}(\theta_i)$ denote Wigner functions of the Euler angles, 
$L$ are the eigenvalues of angular 
momentum, while $\mu$ and $\alpha$ are the eigenvalues of the projections 
of angular momentum on the laboratory fixed $\hat z$-axis and the body-fixed 
$\hat x'$-axis respectively. $\alpha$ has to be an even integer 
\cite{MtVNPA}.

Instead of the projection $\alpha$ of the angular momentum on the 
$\hat x'$-axis, it is customary to introduce the wobbling quantum number 
\cite{MtVNPA,BM} $n_w=L-\alpha$, which labels a series of bands 
with  $L=n_w,n_w+2,n_w+4, \dots$ (with $n_w > 0$) next to the ground state 
band (with $n_w=0$) \cite{MtVNPA}.  

The ``radial'' Eq. (\ref{eq:e42}) is exactly soluble in the case of an 
infinite square well potential ($u(\beta)=0$ for $\beta\leq \beta_W$, 
$u(\beta)=\infty$ for $\beta>\beta_W$). 
Using the transformation $\phi(\beta)=\beta^{-1}f(\beta)$, Eq. (\ref{eq:e42})
becomes a Bessel equation 
\begin{equation}\label{eq:e47}
\Biggl[\frac{\partial^2}{\partial\beta^2} +
\frac{1}{\beta}\frac{\partial}{\partial\beta}
+ \Bigl(\epsilon - \frac{\nu^2}{\beta^2} \Bigr)\Biggr] f(\beta) = 0, 
\end{equation}
with
\begin{equation}\label{eq:e48}
\nu=\sqrt{\lambda+1}=\sqrt{L(L+1) - \frac{3}{4}\,\alpha^2 + 1}
={\sqrt{L(L+4)+3 n_w(2L-n_w)+4}\over 2}.
\end{equation}
Then the boundary condition $f(\beta_W)=0$ determines the spectrum, 
\begin{equation}\label{eq:e10}
\epsilon_{\beta; s,\nu} = \epsilon_{\beta; s,n_w,L} 
= (k_{s,\nu})^2, \qquad k_{s,\nu} = {x_{s,\nu}
\over \beta_W}, 
\end{equation} 
where $x_{s,\nu}$ is the $s$th zero 
of the Bessel function $J_\nu(z)$. The eigenfunctions are  
\begin{equation}\label{eq:e49}
\phi(\beta)=\phi_{s,\nu}(\beta) =\phi_{s,n_w,L}(\beta) 
= \frac{1}{\sqrt{c}}\,\beta^{-1}
J_{\nu}(k_{s,\nu}\beta), \qquad\qquad 
c = \frac{\beta_W^2}{2}\,J^2_{\nu+1}(x_{s,\nu})
\end{equation}
where the normalization constant $c$ is determined from the condition
$\int_0^{\beta_W} \beta^3 \phi^2(\beta) d\beta =1$. 
The notation for the roots has been kept the same as in Ref. 
\cite{IacX5}, while for the energies the notation $E_{s,n_w,L}$ 
will be used. The ground state band corresponds to $s=1$, $n_w=0$.
This model will be called the Z(4) model. 

The calculation of B(E2)s proceeds as in Ref. \cite{Z5}, the only difference 
being that the integrals over $\beta$ have the form 
\begin{equation}\label{eq:e51} 
I_\beta (s_i,L_i,\alpha_i; s_f,L_f,\alpha_f) =
\int_{0}^{\beta_W}\beta\,\phi_{s_i,\nu_i}(\beta)\,\phi_{s_f,\nu_f}(\beta)
\,\beta^3 d\beta,  
\end{equation}
since the volume element in the present case corresponds to four dimensions
instead of five.  

A brief discussion of the interrelations among various triaxial models is now 
in place. In the original triaxial model of Davydov and Filippov 
\cite{DavFil}, the Hamiltonian contains only a rotational term [the second 
term in Eq. (\ref{eq:e41})], and is analytically soluble for all values 
of $\gamma$. In contrast, the Hamiltonian of Davydov and Chaban \cite{DavCha} 
contains both a kinetic energy term [the first term in Eq. (\ref{eq:e41})] and 
a rotational term, and is solved numerically. Meyer-ter-Vehn \cite{MtVNPA}
has shown that triaxial Hamiltonians including both a kinetic energy term and
a rotational term are analytically soluble in the special case of 
$\gamma=30^{\rm o}$. In the present Z(4) case an analytical solution of the 
Davydov and Chaban Hamiltonian is obtained for the special case of 
$\gamma=30^{\rm o}$, as implied by Meyer-ter-Vehn.  

{\bf 3. Relation of the ground state band of Z(4) to E(4)} 

The ground state band of the Z(4) model is related to the second order Casimir 
operator of E(4), the Euclidean group in four dimensions. 
In order to see this, 
one can consider in general the Euclidean algebra in $n$ dimensions, E(n), 
which is the semidirect sum \cite{Wyb} of the algebra T$_n$ of translations 
in $n$ dimensions, generated by the momenta
\begin{equation}\label{eq:e52} 
P_j = -i {\partial \over \partial x_j}, 
\end{equation} 
 and the SO(n) algebra 
of rotations in $n$ dimensions, generated by the angular momenta
\begin{equation}\label{eq:e53} 
 L_{jk} =-i \left(x_j{\partial \over \partial x_k} -x_k {\partial \over
\partial x_j} \right), 
\end{equation}
symbolically written as E(n) = T$_{\rm n}$ $\oplus_s$ SO(n) \cite{Barut}.  
The generators of E(n) satisfy the commutation relations 
\begin{equation}\label{eq:e54} 
 [P_i, P_j] =0, \qquad [P_i, L_{jk}] = i ( \delta_{ik} P_j - \delta_{ij} P_k),
\end{equation}
\begin{equation}\label{eq:e55} 
 [L_{ij}, L_{kl}]=i (\delta_{ik} L_{jl} +\delta_{jl} L_{ik} 
-\delta_{il} L_{jk} -\delta_{jk} L_{il}).
\end{equation}
From these commutation relations one can see that the square of the total 
momentum, $P^2$, is a second order Casimir operator of the algebra, while 
the eigenfunctions of this operator satisfy the equation 
\begin{equation}\label{eq:e56} 
 \left( -{1\over r^{n-1}} {\partial \over \partial r} r^{n-1} {\partial \over 
\partial r} + { \omega(\omega+n-2) \over r^2} \right) F(r) = k^2 F(r), 
\end{equation} 
in the left hand side of which the eigenvalues of the Casimir operator 
of SO(n), $\omega(\omega+n-2)$ appear \cite{Mosh1555}. 
Putting 
\begin{equation}\label{eq:e57} 
 F(r) = r^{(2-n)/2} f(r), 
\end{equation}
and
\begin{equation}\label{eq:e58}
  \nu= \omega+{n-2\over 2},
\end{equation}
Eq. (\ref{eq:e56}) is brought into the form 
\begin{equation}\label{eq:e59} 
 \left( {\partial^2 \over \partial r^2} + {1\over r} {\partial \over \partial 
r} + k^2 - { \nu^2 \over r^2}\right) f(r)  =0, 
\end{equation}
the eigenfunctions of which are the Bessel functions $f(r) = J_\nu(kr) $
\cite{AbrSt}. The similarity between Eqs. (\ref{eq:e59}) and (\ref{eq:e47})
is clear. 

The ground state band of Z(4) is characterized by $n_w=0$, which means 
that $\alpha=L$. Then Eq. (\ref{eq:e48}) leads to $\nu=L/2+1$, while 
Eq. (\ref{eq:e58}) in the case of E(4) gives $\nu=\omega +1$. Then 
the two results coincide for $L=2\omega$, i.e. for even values of $L$. 
One can easily see that this coincidence occurs only in four dimensions. 

{\bf 4. Numerical results and comparisons to E(5) and experiment} 

The lowest bands of the Z(4) model are given in Table 1.
The notation $L_{s,n_w}$ is used. All levels are measured from the ground 
state, $0_{1,0}$, and are normalized to 
the first excited state, $2_{1,0}$. The ground state band is characterized 
by $s=1$, $n_w=0$, while the even and the odd levels of the $\gamma_1$-band 
are characterized by $s=1$, $n_w=2$, and $s=1$, $n_w=1$ respectively, and 
the $\beta_1$-band is characterized by $s=2$, $n_w=0$. These bands are also 
shown in Fig. 1, labelled by $(s,n_w)$. 

Both intraband and interband B(E2) transition rates, normalized to the one 
between the two lowest states, B(E2;$2_{1,0}\to 0_{1,0}$), are given 
in Fig. 1~. 

The similarity between the spectra and B(E2) values of Z(4) and E(5),
for which extensive numerical results can be found in Ref. \cite{E5},  
can be seen in Figs. 2(a) and 2(b), where the spectra of the ground state band 
and the $\beta_1$ band, as well as their intraband B(E2)s are given. 
One can easily check that the similarity extends to interband transitions 
between these bands as well, for which the selection rules in the two models 
are the same. 

The main difference between Z(4) and E(5) appears, as expected, in the 
$\gamma_1$ band, the spectrum of which is shown in Fig. 2(c). The predictions 
of the two models for the odd levels practically coincide, while the 
predictions for the even levels differ, since in the E(5) model the levels 
are exactly paired as (3,4), (5,6), (7,8), \dots, as imposed by the underlying 
SO(5)$\supset$SO(3) symmetry \cite{IacE5,E5}, while in the Z(4) model
the levels are approximately paired as (4,5), (6,7), (8,9), \dots, which 
is a hallmark of rigid triaxial models \cite{DavFil}. The latter behaviour 
is never materialized fully \cite{ZC}, but it is known \cite{Casten} that 
$\gamma$-unstable models and $\gamma$-rigid models yield similar predictions 
for most observables if $\gamma_{rms}$ of the former equals $\gamma_{rigid}$ 
of the latter, a situation occuring in the Ru-Pd, Xe-Ba (below $N=82$), 
and Os-Pt regions. 

Predictions of the Z(4) model are compared to existing experimental data for 
$^{128}$Xe \cite{Xe128}, $^{130}$Xe \cite{Xe130}, and $^{132}$Xe \cite{Xe132}
in Fig. 3. The reasonable agreement observed is in no contradiction with 
the characterization of these nuclei as O(6) nuclei \cite{Casten}, since, 
as mentioned above, the predictions of $\gamma$-unstable models (like O(6) 
\cite{IA}) and $\gamma$-rigid models (like Z(4)) for most observables are 
similar if $\gamma_{rms}$ of the former equals $\gamma_{rigid}$ of the latter. 
    
{\bf 5. Discussion}

In the present work an exact solution of the Bohr Hamiltonian with 
$\gamma$ ``frozen'' to $30^{\rm o}$, called Z(4), is obtained. Spectra 
and B(E2) transition rates of Z(4) resemble these of the critical point 
symmetry E(5), while the ground state band of Z(4) is related to the 
Euclidean algebra E(4), thus offering a first clue of connection between 
critical point symmetries and Lie algebras, in addition to the case 
of the E(5) model \cite{IacE5}. Empirical evidence for Z(4) 
in the Xe region around $A=130$ has been presented. 

It should be emphasized, however, that neither the similarity of spectra 
and B(E2) values of Z(4) to these of the E(5) model, nor the coincidence
of the ground state band of Z(4) to the spectrum of the Casimir operator
of the Euclidean algebra E(4)
clarify the algebraic structure of the Z(4) model, the symmetry algebra of 
which has to be constructed explicitly, starting from the fact that $\gamma$ 
is fixed to $30^{\rm o}$. The fact that the Bohr Hamiltonian for 
$\gamma=30^{\rm o}$ possesses ``accidentally'' a symmetry axis (the 
body-fixed $\hat x'$-axis) has been early realized \cite{Brink}.    
This ``accidental'' symmetry should also serve as the starting point for 
clarifying the symmetry underlying other solutions of the Bohr Hamiltonian 
obtained for $\gamma=30^{\rm o}$ \cite{Z5,Jolos,Fortunato}. 

{\bf Acknowledgements} 

One of the authors (IY) is thankful to the Turkish Atomic Energy Authority 
(TAEK) for support under project number 04K120100-4.


\begin{table}

\caption{Energy levels of the Z(4) model, measured from the 
$L_{s,n_w}=0_{1,0}$ ground state and normalized to the $2_{1,0}$ lowest 
excited state. 
}

\bigskip

\begin{tabular}{ r r r r | r r}
\hline
$s,n_w$ & 1,0 & 1,2 & 2,0  &   & 1,1   \\
$L$&       &       &       &$L$&       \\ 
\hline
   &       &       &       &   &       \\
 0 & 0.000 &       & 2.954 &   &       \\
 2 & 1.000 & 1.766 & 4.804 & 3 & 2.445 \\
 4 & 2.226 & 4.051 & 6.893 & 5 & 4.239 \\
 6 & 3.669 & 6.357 & 9.215 & 7 & 6.188 \\
 8 & 5.324 & 8.788 &11.765 & 9 & 8.316 \\
10 & 7.188 &11.378 &14.538 &11 &10.630 \\
12 & 9.256 &14.139 &17.531 &13 &13.135 \\
14 &11.526 &17.079 &20.742 &15 &15.831 \\
16 &13.996 &20.202 &24.167 &17 &18.719 \\
18 &16.665 &23.509 &27.805 &19 &21.799 \\
20 &19.530 &27.003 &31.653 &21 &25.071 \\
\hline
\end{tabular}
\end{table}


\begin{figure}[ht] 
{\includegraphics[height=170mm]{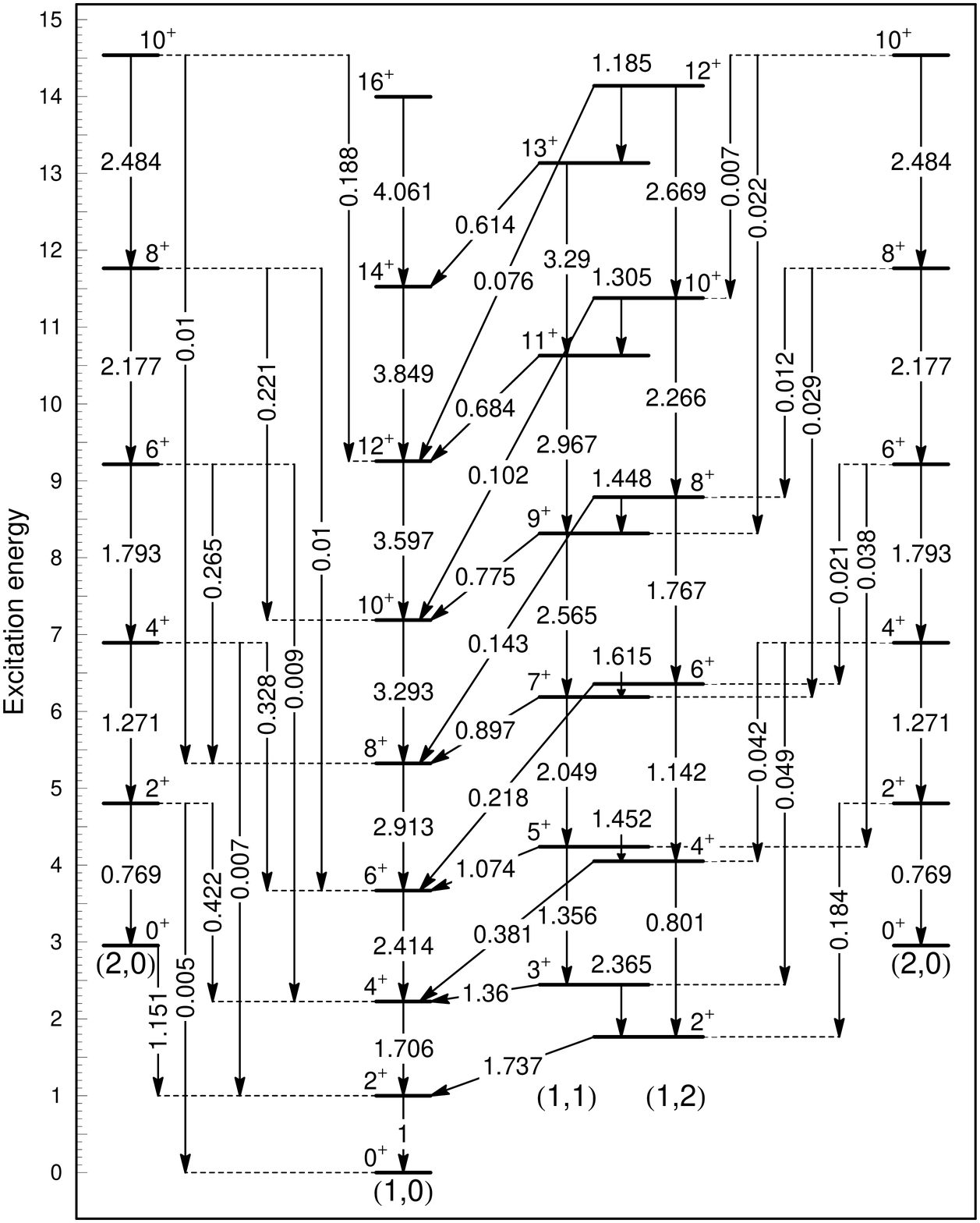}} 
\caption{Intraband and interband B(E2) transition rates in the Z(4) 
model, normalized to the B(E2;$2_{1,0}\to 0_{1,0}$) rate. Bands are 
labelled by $(s,n_w)$, their levels being normalized to $2_{1,0}$.  
The (2,0) band is shown both at the left and at 
the right end of the figure for drawing purposes.}  
\end{figure}


\begin{figure}[ht]
\rotatebox{270}{\includegraphics[height=85mm]{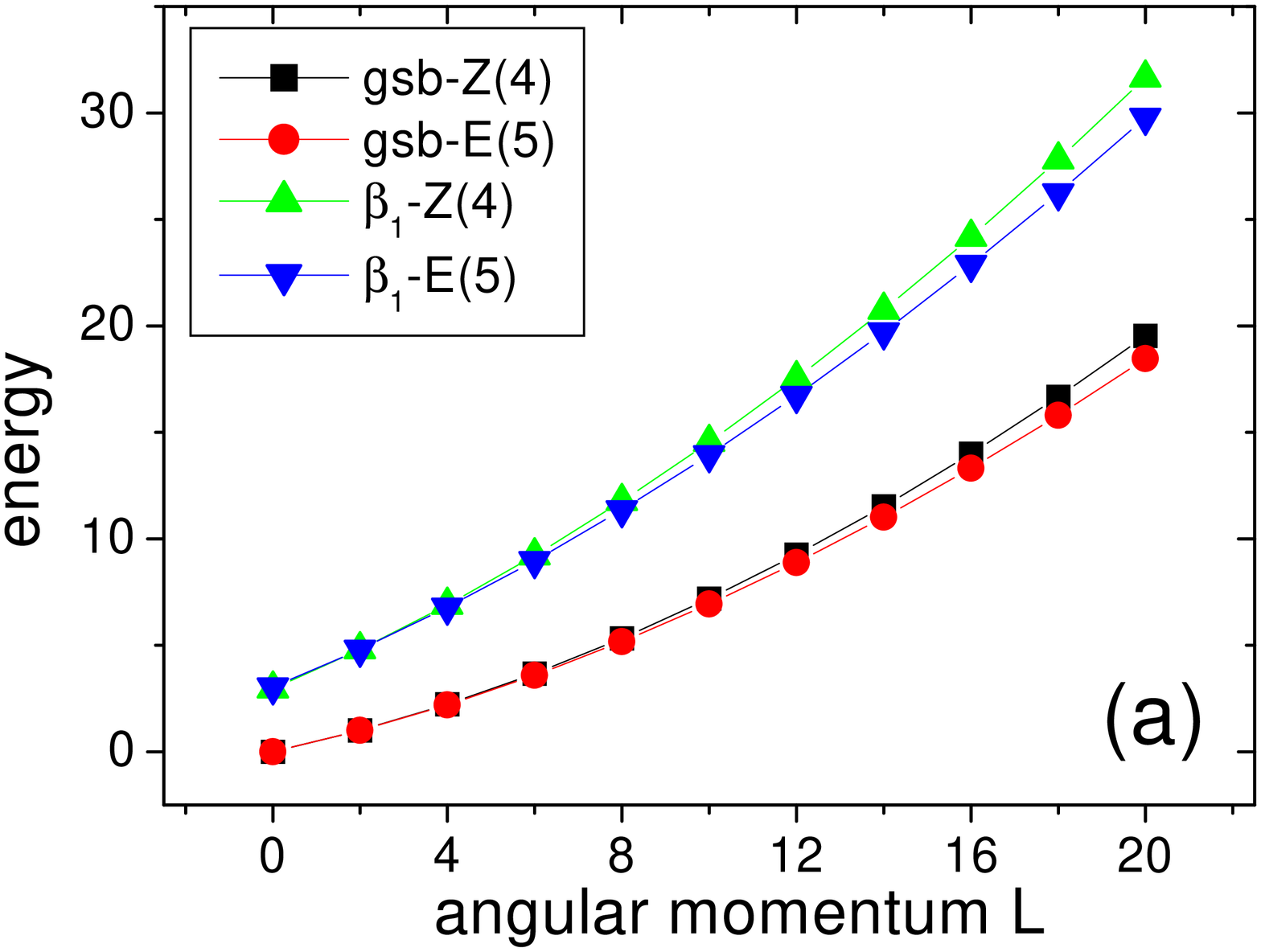}} 
\rotatebox{270}{\includegraphics[height=85mm]{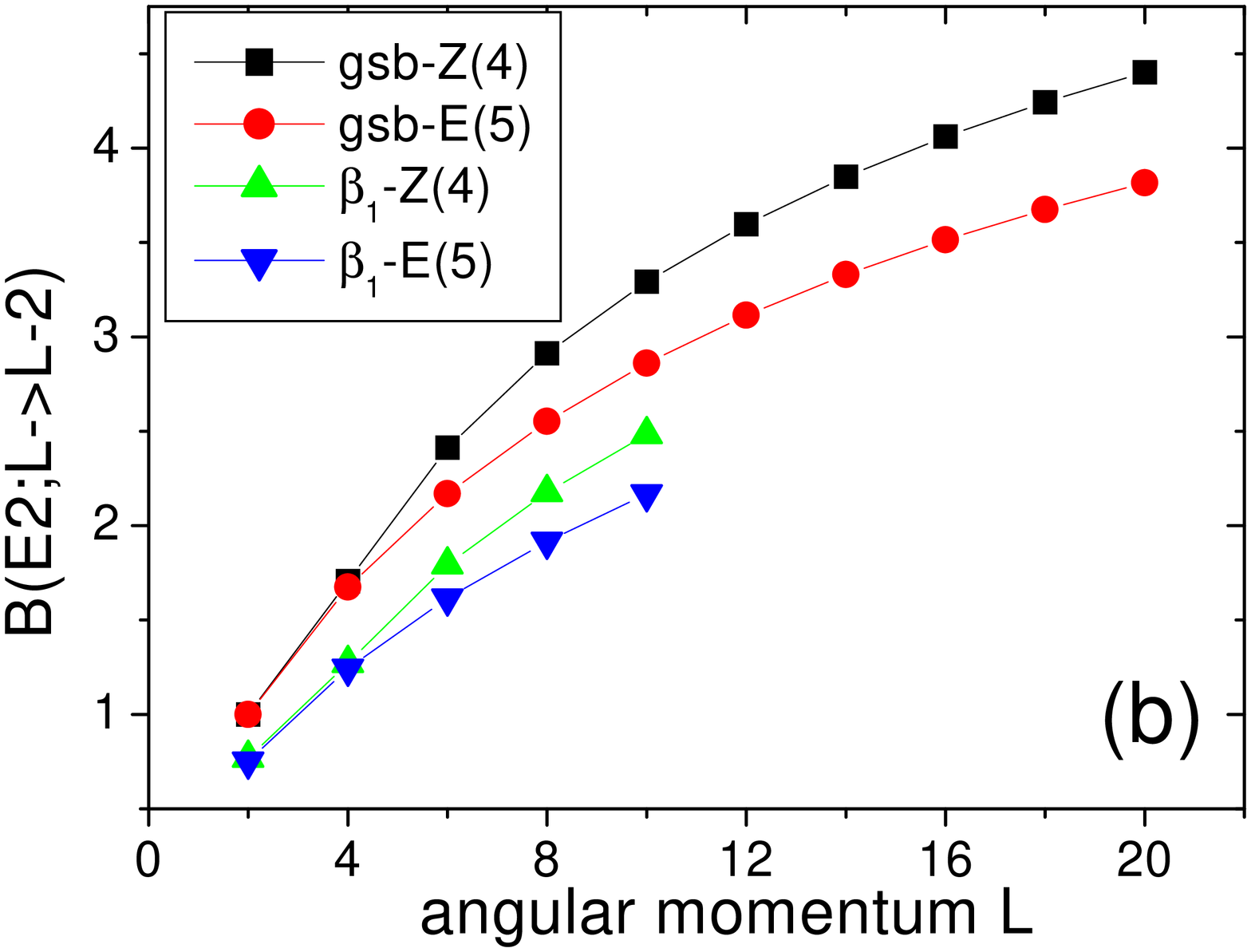}}
\rotatebox{270}{\includegraphics[height=85mm]{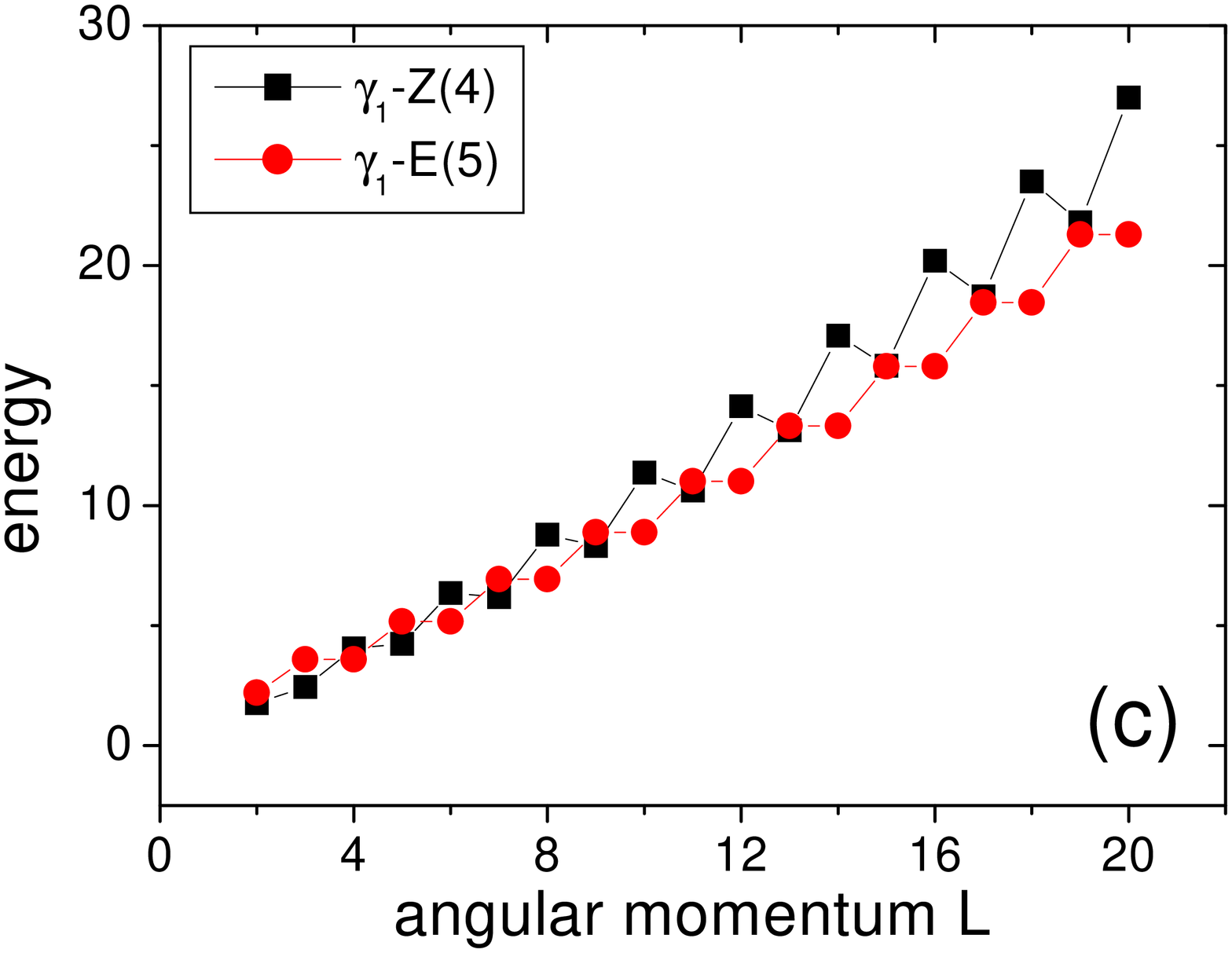}}
\caption{(a) Ground state band [$(s,n_w)=(1,0)$] and first excited band 
[$(s,n_w)=(2,0)$] of Z(4) (labeled as $\beta_1$-band)
compared to the corresponding bands of E(5) \cite{IacE5,E5}. In each model 
all levels are normalized to the $2_1^+$ state. 
(b) Intraband B(E2) transition rates within the same bands of Z(4) 
compared to the corresponding B(E2) rates of E(5). In each model all rates 
are normalized to the $2_1^+\to 0_1^+$ rate. 
(c) The lowest``$K=2$ band'' of Z(4) [formed out of the ($s,n_w$) bands (1,2) 
and (1,1), labeled as $\gamma_1$], compared to the corresponding band of E(5).}
\end{figure}


\begin{figure}[ht]
{\includegraphics[height=95mm]{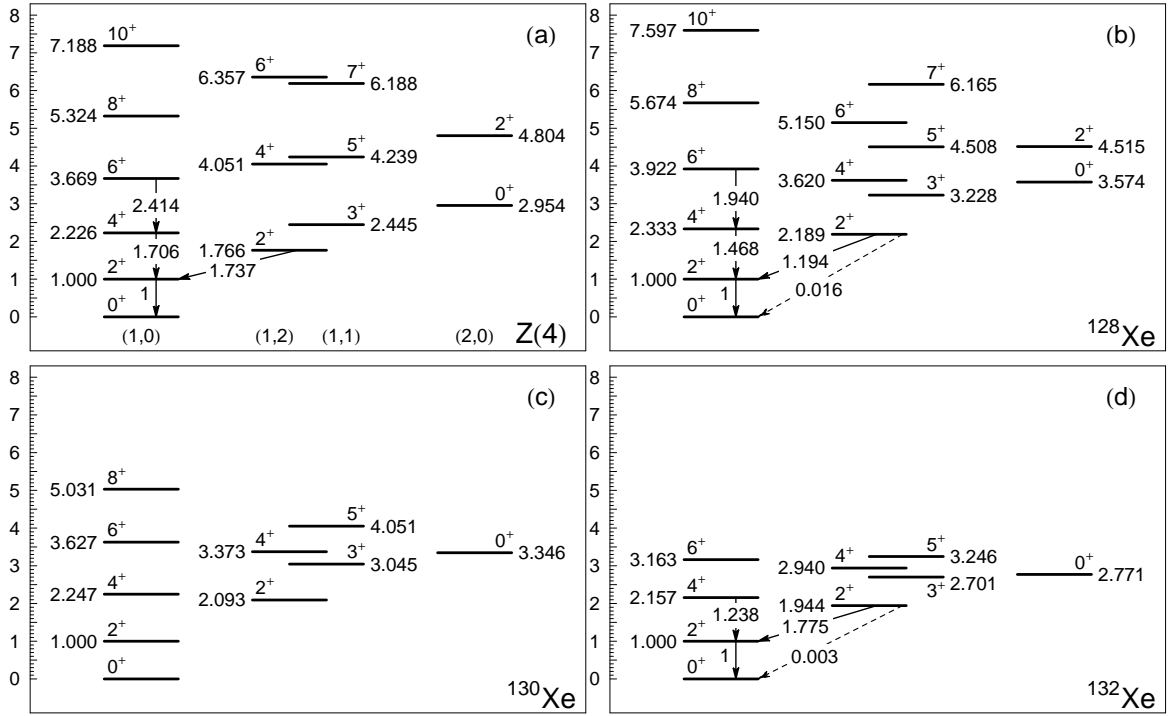}} 
\caption{Comparison of the Z(4) predictions for (normalized) energy levels 
and (normalized) B(E2) transition rates (a) to experimental data for 
$^{128}$Xe \cite{Xe128} (b), $^{130}$Xe \cite{Xe130} (c), 
and $^{132}$Xe \cite{Xe132} (d). Bands in (a) are labelled by $(s,n_w)$.  
See Section 4 for further discussion.} 
\end{figure}

\end{document}